\documentclass[11pt,a4paper]{article}
\usepackage[cp1251]{inputenc}
\usepackage{indentfirst}
\usepackage{amsmath,amsthm,amssymb}
\usepackage{graphicx}
\usepackage{setspace}

\begin{document}

\textwidth=135mm
 \textheight=200mm
\begin{center}
{\bfseries Weyl group, CP  and the kink-like field configurations in the effective   $SU(3)$ gauge theory}
\vskip 5mm B.V. Galilo $^{\dag\ddag}$, S.N. Nedelko $^{\dag}$ 

{\small {\it $^\dag$ BLTP, Joint Institute for Nuclear Research, 141980, Dubna, Russia}} \\
{\small {\it $^\ddag$ Department of Theoretical Physics,
International University  ''Dubna'',
141980, Dubna, Russia}}
\\
\end{center}


\begin{abstract}
 Effective Lagrangian for  Yang-Mills gauge fields invariant under the standard space-time  and  local gauge  $SU(3)$  transformations is considered. It is demonstrated that a set of twelve  degenerated  minima  exists as soon as a nonzero gluon condensate is postulated. The minima are connected to each other by the parity transformations and Weyl group transformations associated with the color $su(3)$ algebra.  The presence of degenerated discrete minima in the effective potential leads to the solutions of the effective Euclidean equations of motion in the form of the kink-like gauge field configurations interpolating between different minima.  Spectrum of charged scalar field in the kink background  is discussed.\\

\noindent
PACS numbers: 12.38.Aw, 12.38.Lg, 14.70.Dj

\end{abstract}

\section{Introduction}

The purpose of the paper is to expose potentially interesting relation of the Weyl group  associated with the color gauge  $SU(3)$  symmetry to the structure of 
QCD  vacuum. At the level of classical Yang-Mills Lagrangian the Weyl group symmetry is trivial.  However, the vacuum structure of QCD is determined by quantum effects. 
The  standard way to discuss the vacuum structure of the theory in terms of effective quantum action   relates to  the Landau-Ginsburg type construction  based on the symmetries of the theory.  

We consider  the Landau-Ginsburg Lagrangian for pure Yang-Mills gauge fields invariant under the standard space-time  and  local gauge  $SU(3)$  transformations. It is demonstrated that a set of twelve  degenerated  minima of the action density exists as soon as a nonzero gluon condensate is postulated in the action. The minima are connected to each other by the Weyl group transformations associated with the color $su(3)$ algebra and  parity transformation. The presence of degenerated discrete minima in the Lagrangian leads to the solutions of the effective equations of motion in the form of the kink-like gauge field configurations interpolating between different minima. As an example we write down the simplest solution which interpolates between self-dual and anti-self-dual Abelian homogeneous fields and consider the spectrum of covariant derivative squared $D^2$  in the presence of this kink background field.   The kink configuration is seen here as a domain wall separating the regions with almost self-dual and anti-self-dual abelian gauge field.  It should be stressed that consideration itself and the results of this paper  can be seen as  instructive but very preliminary ones.  

\section{Motivation}

There are several observations and phenomenological estimates which can be considered  as a qualitative motivation for  introducing the effective action to be discussed in the next section.

 The Weyl group associated with  $SU(N)$ gauge theory  can be conveniently exposed in terms of  representation of the gauge fields suggested in a series of
  papers by   S. Shabanov~\cite{shabanov1, shabanov2}, Y.M.~Cho \cite{Cho1,Cho2},  L.D.~Faddeev and  A. J. Niemi \cite{Faddeev} and, recently, by K.-I. Kondo \cite{Kondo}. In this parameterization the Abelian part ${\hat V}_\mu (x)$ of the gauge field ${\hat A}_\mu (x)$ is
 separated manifestly,
\begin{eqnarray}
\label{eqnsFaddeevNiemKondo} 
 {\hat A}_\mu (x) &=& {\hat V}_\mu (x) + {\hat X}_\mu (x), \, 
{\hat V}_\mu (x) = {\hat B}_\mu (x) + {\hat C}_\mu (x), \\
 {\hat B}_\mu (x) &=& [n^aA^a_\mu (x)]\hat{n} (x)=B_\mu(x)\hat{n}(x), \nonumber \\
 {\hat C}_\mu (x) &=& g^{-1}\partial_\mu \hat{n}(x)\times \hat{n}(x), \nonumber\\
 {\hat X}_\mu (x) &=& g^{-1}{\hat n}(x) \times \left( \partial_\mu {\hat n}(x) + g {\hat A}_\mu (x) \times {\hat n}(x) \right), \nonumber
\end{eqnarray}
where ${\hat A}_\mu (x) = A^a_\mu (x) t^a$, ${\hat n} (x) = n_a (x) t^a$,   $n^a n^a = 1$, and
\begin{eqnarray*}
{\partial_\mu\hat n}\times {\hat n} = i f^{abc}\partial_\mu n^a n^b t^c,\,
      \, [t^a,t^b]=if^{abc}t^c.
\end{eqnarray*}
The field ${\hat V}_\mu $ is seen as the Abelian field in the sense that $[{\hat V}_\mu (x),{\hat V}_\nu (x)]=0$.

The comprehensive analysis of the RG-improved one-loop effective action for the Abelian component $ {\hat B}_\mu (x)$ with the constant $n^a$ and the covariantly constant (anti)self-dual field  $B_\mu=-(1/2) B_{\mu\alpha} x_\alpha$ was given long time ago by Minkowski~\cite{Minkowski},  Pagels and Tomboulis~\cite{Pagels}, and Leutwyler~\cite{Leutwyler}.
The analysis based on  the trace anomaly of the energy-momentum tensor and  renormalization group  leads to the following form of the effective potential for the $SU(2)$ gauge group~\cite{Pagels}:
\begin{eqnarray}
\label{LeffRG}
U_{\rm eff}^{\rm RG}=B^2 \left[ \frac{1}{g^2 (\lambda B / \Lambda^2)} + \varepsilon_0  \right].
\end{eqnarray}
Here constant  $B$ is defined as $B_{\mu\alpha}B_{\nu\alpha}=B^2\delta_{\mu\alpha}$. For the strong field $B\gg \Lambda$ this expression agrees with the result of the explicit one-loop calculation~\cite{Leutwyler}
\begin{eqnarray}
\label{Leff1loop}
U_{\rm eff}^{\rm 1-loop}=B^2 \left[ \frac{11}{24\pi^2}\ln\frac{\lambda B}{  \Lambda^2} + \varepsilon_0  \right].
\end{eqnarray}
Equation (\ref{LeffRG}) indicates that the effective potential can have a minimum for nonzero strength $B$ only for the negative value of the parameter $\varepsilon_0$. This parameter can be treated as the dielectric constant as $g\to\infty$. One loop result (\ref{Leff1loop}) displays the strong field asymptotics  $B^2\ln( B/  \Lambda^2)$ of the effective Lagrangian and, hence, its boundedness from below.  In both calculations the constant $\varepsilon_0$ is a free parameter. Some knowledge about the sign of  $\varepsilon_0$ could be obtained from the lattice calculations. A minimum of the effective Lagrangian at nonzero field strength was reported in~\cite{Trottier}. However the most interesting region of small field strength is the most difficult one for the lattice calculation, and, as the authors of~\cite{Trottier}  stressed, this result should be taken into account with great care. Existence of  nonzero gluon condensate can be considered as a general phenomenological argument in favor of nonzero value of the field strength at the minimum of the QCD effective action and, hence, the negative value of  $\varepsilon_0$. Certainly, the ordered state corresponding to a plain constant field cannot be  considered as an appropriate approximation for QCD vacuum  as it breaks  all  the symmetries of QCD at once.  Required disorder in the mean field could be provided  by an ensemble of gauge field configurations with the  strength being constant almost everywhere but changing directions in space and color space as well as self-  and anti-self-duality  in  small regions of space-time reminiscent of the domain walls. Phenomenological model of confinement, chiral symmetry breaking and hadronization based on the ensemble of Abelian (anti-)self-dual fields was developed in a series of papers~\cite{EN,NK1,NK2}. 
The dominance of the domain structured gauge field configurations has been observed in the recent lattice calculations~\cite{Moran:2008xq, Moran:2007nc,deForcrand:2008aw,deForcrand:2006my,Ilgenfritz:2007ua}. In paper~\cite{deForcrand:2008aw} an effective model of $SU(2)$ gauge theory for the domain wall formation was considered.  The center symmetry realization in lattice version of QCD is in the focus of these studies. 	In this paper, we  consider the effective Lagrangian which displays another  model for the domain wall formation based  on CP and  the Weyl symmetry  breakdown triggered by the trace anomaly or, equivalently, the nonzero gluon condensate.

\section{Effective Lagrangian}

Consider the following  effective Lagrangian  for the  gauge fields satisfying the requirements of invariance under the gauge group $SU(3)$ and space-time  transformations,
\begin{eqnarray}
\label{Leff}
    L_{\rm eff} &=& -\frac{1}{4}\left(D^{ab}_\nu F^b_{\rho\mu} D^{ac}_\nu F^c_{\rho\mu} + D^{ab}_\mu F^b_{\mu\nu} D^{ac}_\rho F^c_{\rho\nu }\right)
     - U_{\rm eff} \\
U_{\rm eff}&=&\frac{1}{12} {\rm Tr}\left(C_1\hat{F}^2 + \frac{4}{3}C_2\hat{F}^4 - \frac{16}{9}C_3\hat{F}^6\right),
\nonumber
\end{eqnarray}
where
\begin{eqnarray*}
 && D^{ab}_\mu = \delta^{ab} \partial_\mu - i\hat{A}^{ab}_\mu = \partial_\mu - iA^c_\mu {(T^c)^{ab}},
\\
 && F^a_{\mu\nu} = \partial_\mu A^a_\nu - \partial_\nu A^a_\mu - if^{abc} A^b_\mu A^c_\nu,
\\
 && \hat{F}_{\mu\nu} = F^a_{\mu\nu} T^a,\ \ \ T^a_{bc} = -if^{abc}
\\
 && {\rm Tr}\left(\hat{F}^2\right) = \hat{F}^{ab}_{\mu\nu}\hat{F}^{ba}_{\nu\mu} = -3 F^a_{\mu\nu}F^a_{\mu\nu} \leq 0,
\\ && C_1>0, \ C_2>0, \ C_3 > 0.
\end{eqnarray*}
The gauge coupling constant is absorbed into the  gauge field, $gA_\mu\to A_\mu$. The signs of the constants $C_1$, $C_2$ and $C_3$ are chosen in such a way that the effective Lagrangian is bounded from below and has a minimum at nonzero
value of the field strength squared
\begin{eqnarray}
&&  F^a_{\mu\nu} F^a_{\mu\nu} = 4 b_{\rm vac}^2\Lambda^4 > 0.
\nonumber\\
&& b^2_{\rm vac}=
\left(\sqrt{C_2^2+3C_1C_3}-C_2\right)/3C_3.
\nonumber
\end{eqnarray}
 In terms of Eq.~(\ref{LeffRG}) the choice of sign of $C_1$ corresponds to the negative $\varepsilon_0$.  Lagrangian (\ref{Leff}) contains the lowest order covariant derivatives and the effective potential  which has a polynomial in $F^2$ form. Thus the field ${\rm Tr} F^2$ plays the role of the order parameter.  The presence of the term $\hat{F}^6$ is of the crucial importance since the Weyl group becomes manifest  only  in  this and higher orders in field strength. Further increase in the polynomial order  in (\ref{Leff}) does not change qualitatively the character of the Weyl group realization. The form of Effective Lagrangian (\ref{Leff}) is not the most general one. Our aim is to study an instructive example rather than to deal with the full problem in all its complexity. Namely, let us  consider a set of  fields $A_\mu$   with the Abelian field strength of the following form
\begin{eqnarray*}
\label{B&A_relationship}  && \hat{F}_{\mu\nu}  = \hat{n}B_{\mu\nu},
\end{eqnarray*}
where  matrix  $\hat{n}$ is an element of Cartan subalgebra in the adjoint representation
\begin{eqnarray*}
\label{hatn}
 \hat{n} = T^3\ \cos\left(\xi\right) + T^8\ \sin\left(\xi\right), 
\, 
0 \leq \xi < 2\pi.
\end{eqnarray*}
For $\xi={\rm const}$ this field corresponds to the Abelian part  ${\hat B}_\mu (x)$ of the gauge field in the representation  (\ref{eqnsFaddeevNiemKondo}). It is convenient to introduce the following notation:
\begin{eqnarray*} 
\label{notation_beh}
&& \hat{b}_{\mu\nu} =\hat{n}B_{\mu\nu}/\Lambda^2 = \hat{n}b_{\mu\nu}, \ 
b_{\mu\nu}b_{\mu\nu}=4b_{\rm vac}^2,
\\
 &&   e_i=b_{4i},\ 
    h_i = \frac{1}{2}\varepsilon_{ijk}b_{jk},\ 
    \left(\mathbf{eh}\right) = \left|\mathbf{e}\right| \left|\mathbf{h}\right| \cos{\omega},
\\
 &&   \mathbf{e}^2 + \mathbf{h}^2 = 2b_{\rm vac}^2,\ 
 \left(\mathbf{eh}\right)^2 = \mathbf{h}^2\left(2b^2-\mathbf{h}^2\right) \cos^2
 {\omega}.
\end{eqnarray*}
With this notation one arrives at the following formulae for traces:
\begin{eqnarray*}
\label{traces}
   {\rm Tr}\hat{b}^2 = -12b_{\rm vac}^2,\ 
   {\rm Tr}\hat{b}^4 = 18\left(b_{\rm vac}^4-\frac{1}{2}
   \left(\mathbf{eh}\right)^2\right),\ 
   {\rm Tr}\hat{b}^6 = -3b_{\rm vac}^2\left(10 + \cos6\xi\right) \left(b_{\rm vac}^4 - \frac{3}{4}{\left(\mathbf{eh}\right)}^2\right).
\end{eqnarray*}
Respectively, the effective potential takes the form
\begin{eqnarray}
  \label{Ueff(b)}
 U_{\rm eff} &=& \Lambda^4\left[-C_1b_{\rm vac}^2+C_2\left(2b_{\rm vac}^4 - \left(\mathbf{eh}\right)^2\right) + \frac{1}{9}C_3b^2\left(10 + \cos6\xi\right) \left(4b_{\rm vac}^4 - 3\left(\mathbf{eh}\right)^2\right)\right].
\end{eqnarray}
The potential (\ref{Ueff(b)}) is invariant under transformations
$\xi \rightarrow \xi +\pi k/3$, $k = 1,\dots,6 $, which
can be seen as specific rotations of $\hat{n}$ in Cartan subalgebra. These transformations lead to permutations of the eigenvalues of  $\hat n$ and, hence, do not change the traces of  $\hat n^{k}$. The permutations correspond to the Weyl group associated with $su(3)$ algebra, i.e., the group of reflections of the roots of $su(3)$.  The
effective potential is invariant with respect to parity transformation which results in the degeneracy of the self- and anti-self-dual fields corresponding to $\omega=0,\pi$. Altogether there are twelve discrete global degenerated minima at the following values of the variables $h$, $\omega$ and $\xi$
\begin{equation}
\label{minima}
 \mathbf{h}^2 = b_{\rm vac}^2>0, \ \   \omega=\pi k \ \ (k=0,1),    \ \  \xi_n = \frac{\pi}{6}\left(2n+1\right) (n = 0,\dots, 5).
\end{equation}
It should be stressed here that we have postulated in (\ref{Leff}) the minimum at nonzero value of the scalar gauge invariant field $F^a_{\mu\nu} F^a_{\mu\nu}$ equivalent to the existence of nonzero gluon condensate, but the set of minima in $\omega$ and $\xi$  appeared  as a consequence of the space-time and local gauge symmetries of the effective Lagrangian (\ref{Leff}). Inclusion of higher powers of $\hat F$ does not change this picture qualitatively, but the presence of the term $o(\hat F^6)$ is crucial since dependence on $\xi$ appears starting the 6th order in $\hat F$. The minimum of the effective potential in $\xi$ is achieved for the values $\xi_n$ corresponding to the boundaries of the  Weyl chambers in the root space of $su(3)$. Existence of the degenerate minima in the effective action related to the Weyl group was reported earlier in \cite{NK1} and, in the context of the one-loop effective potential of $SU(N)$ gauge theory  in \cite{Cho2}.  A mechanism of gauge field localization on a domain wall within the framework of one-loop effective action for  pure Yang-Mills theory was presented in \cite{kobakhidze}.

\section{Kink-like configurations}

It is well-known that the presence of the discrete global vacua in a system leads  to  the existence of  kink-like solutions of the equations of motion. These solutions describe field configurations interpolating between different vacua and can be treated as the domain walls between regions in $R^4$ with particular constant values of the parameters $\omega$ and $\xi$ from the set (\ref{minima}). In order to exemplify this statement, let us write down Lagrangian in terms of the fields $\xi(x)$, $\omega(x)$ and $\mathbf{h}(x)$.

 Suppose $\mathbf{e}^2 (x)\equiv \mathbf{h}^2 (x)\equiv b_{\rm vac}^2$.
Then we have
\begin{eqnarray*}
  &&\delta U_{\rm eff}= U_{\rm eff}-U_{\rm eff}^{\rm min} = b_{\rm vac}^4  \Lambda^4 \left[(C_2 + 3 C_3 b_{\rm vac}^2 ){\sin^2{\omega}} + \frac{1}{9}C_3b_{\rm vac}^2  (10+\cos6\xi ) \left(1+3\sin^2{\omega}\right)\right],
\nonumber
\\
  && \frac{1}{4} \partial_\mu F^a_{\rho\sigma} \partial_\mu F^a_{\rho\sigma} = \frac{\Lambda^2}{2}\left(\partial_\mu \mathbf{ h} \partial_\mu \mathbf{h} + \partial_\mu \mathbf{e} \partial_\mu\mathbf{e} + b_{\rm vac}^2 \ \partial_\mu \xi\ \partial_\mu \xi \right).
\end{eqnarray*}
Here $U_{\rm eff}^{\rm min}$ is the minimal value of the effective potential  corresponding to  the constant values of  $h$, $\omega$ and $\xi$ given in (\ref{minima}).

 \begin{figure}
 \centerline{   \includegraphics[width=70mm,angle=0]{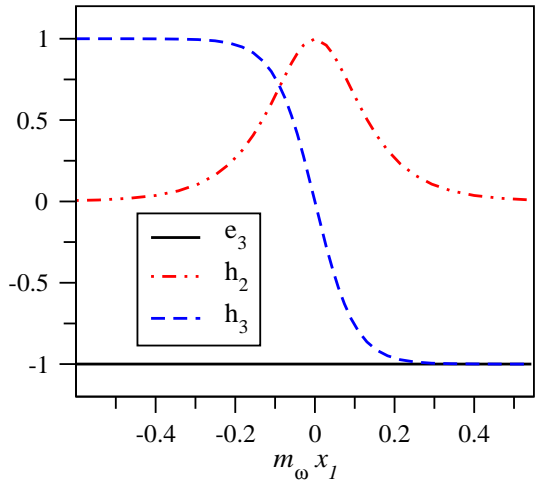}}
    \caption{The gauge field flips from the anti-self-dual at $m_\omega x_1\ll-1$ to self-dual at $m_\omega x_1\gg 1$ configuration: $h_3=b_{\rm vac}\cos\omega$, $h_2=b_{\rm vac}\sin\omega$, $e_i=\delta_{i3}b_{\rm vac}$.  Here $b_{\rm vac} = 1$,  $m_\omega=10\Lambda$.}
\label{comp}
 \end{figure}

In order to separate the relevant variable $\omega$ 
from other degrees of freedom of {\bf e} and  {\bf h}
it is convenient to represent the electric field as
\begin{eqnarray*}
\label{rot}
&&e_i(x)=O_{ij}(x) h_j(x),\nonumber\\
&&O_{ij}=\delta_{ij}\cos\omega(x)+m_i(x)m_j(x)(1-\cos\omega(x))
+\varepsilon_{ijk}m_k(x)\sin\omega(x)
\end{eqnarray*}
where $O$ is a local rotation about unit vector {\bf m} orthogonal to {\bf h},
\begin{eqnarray*}
m_i(x)=\frac{1}{b^2_{\rm vac}\sin\theta(x)}
\left(b^2_{\rm vac}\delta_{ij}-h_i(x)h_j(x)\right)v_j.
\end{eqnarray*}
Here $\theta$ is azimuthal angle of {\bf h} with respect to $\bf v$, 
and {\bf v} is a constant unit vector. We will take
$$
v_i=\delta_{i3}, \ \ m_i=(b_{\rm vac}\delta_{i3}-h_i\cos\theta)/b_{\rm vac}
\sin\theta.
$$
After some algebraic transformations the Lagrangian density $L_{\rm eff}$  takes the form
\begin{eqnarray*}
    L_{\rm eff}
    &=& - \frac{1 }{2}\Lambda^2 b_{\rm vac}^2 
        \left(\partial_\mu \xi \partial_\mu \xi + \partial_\mu \omega \partial_\mu \omega
                + \left(\cos^2{\omega} + \frac{\sin^2{\omega}}{\sin^2{\theta}}\right)
                \partial_\mu \theta \partial_\mu \theta
    + \right.
    \nonumber
    \\
    &&\left.
    +\ \sin^2 \theta \left(1+\cos^2 \omega\right)\partial_\mu \varphi \partial_\mu \varphi
    + 2{\cos^2\omega}\sin\theta \partial_\mu \omega \partial_\mu \varphi - \sin{2\omega} \cos{\theta} \partial_\mu
    \theta \partial_\mu \varphi \right)
    -
    \nonumber\\
    && -\ b_{\rm vac}^4 \Lambda^4\left(\left(C_2+3C_3b_{\rm vac}^2 \right) \sin^2{\omega}+\frac{1}{9}C_3b_{\rm vac}^2 \left(10+\cos{6\xi}\right)\left(1+3\sin^2{\omega}\right)\right).
\end{eqnarray*}
Let $\cos\left(6\xi \right) = -1$, $\theta = \rm const$ and $\varphi =
\rm const$, then
\begin{eqnarray*}
    L_{\rm eff} = -\frac{1 }{2}\Lambda^2 b_{\rm vac}^2 \partial_\mu \omega \partial_\mu \omega - b_{\rm vac}^4  \Lambda^4 \left(C_2+3C_3b_{\rm vac}^2 \right){\sin^2\omega},
\end{eqnarray*}
and the Euler-Lagrange equation 
\begin{eqnarray*}\label{sinGordonextended}
    \partial^2\omega = m_\omega^2 \sin 2\omega,\ \   m_\omega^2 = b_{\rm vac}^2  \Lambda^2\left(C_2+3C_3b_{\rm vac}^2 \right).
\end{eqnarray*}
Let us look for solutions $\omega$ which depend only  on one of the coordinates, say $x_1$. Equation (\ref{sinGordonextended}) takes the form of sin-Gordon equation
\begin{eqnarray*}\label{sinGordon}
   \omega''(x_1) = m^2_{\omega} \sin 2\omega(x_1),
\end{eqnarray*}
with kink  solution
\begin{eqnarray}\label{kink}
    \omega = 2\ {\rm arctan}\left(\exp\left(\sqrt{2}m_{\omega} x_1\right)\right).
\end{eqnarray}
According to Eq.(\ref{kink}), the angle between chromoelectric and chromomagnetic fields $\omega (x_1)$ varies from $\pi$ to $0$  for   $x_1\in [-\infty,\infty]$. It corresponds to the change from anti-self-dual to self-dual gauge field configuration, as is shown in Fig.\ref{comp}

Similarly, if  $\sin(\omega) =0$, $\theta = \rm const$ and $\varphi =
\rm const$, then 
\begin{equation*}
   \partial^2\xi = -\frac{1}{3}m_\xi^2 \sin6\xi,\ \   m^2_{\xi} = 2C_3\Lambda^2b_{\rm vac}^4,
\end{equation*}
with the solution
\begin{eqnarray*}
 \xi_k(x_1) =\frac{1}{3} \arctan\left[\sinh(m_\xi x_1)\right]+\frac{\pi k}{3}, \ k=1,\dots,6,
\end{eqnarray*}
interpolating between two consequent vacuum values of $\xi$ in (\ref{minima}) associated with the boundaries of the $k$-th Weyl chamber.

\section{Spectrum of the charged field in the kink-like background}

In this section, we estimate the change in the spectrum of  color charged scalar field caused by the kink-like defect in $\omega$ in comparison with  the spectrum in 
the presence of confining   (anti-)self-dual purely homogeneous abelian background. Here we consider the infinitely thin domain wall for $\omega$ which corresponds to  $m_\omega\gg\Lambda$ in Eq.~(\ref{kink}).
  Since the kink interpolates between the CP conjugated vacua and some particular vacuum value of angle $\xi$ it is sufficient to consider the eigenvalue problem 
\begin{eqnarray}
\label{eigeneqn}
- \left( \partial_\mu - iB_\mu(x)\right)^2\phi =
\lambda\phi.
\end{eqnarray}
In the case of infinitely thin wall, the field $B_\mu(x)$ is self-dual for $x_1<0$, anti-self-dual for  $x_1>0$, but inside the domain wall (at $x_1=0$) electric and magnetic fields are orthogonal to each other. 

Inside the domain bulk vector potential can be represented as the homogeneous  self- or anti-self-dual field
\begin{eqnarray*}
\label{field_B}
 &&B_{\mu}(x) = B_{\mu\nu}x_\nu,\ \ \tilde B_{\mu\nu}=\pm B_{\mu\nu}, \ B_{\mu\alpha}B_{\nu\alpha}=B^2\delta_{\mu\nu}, \ B=\Lambda^2 b_{\rm vac}.
\end{eqnarray*}
Squire integrable solution is well known in this case. The following field strength configuration can be chosen without loss of generality
\begin{eqnarray*}\label{fields_HE}
 H_1 = H_2 = 0,\  H_3 = \mp 2B,\ \ \ E_1 = E_2 = 0,\ 
 E_3 = -2B.
\end{eqnarray*}
Equation (\ref{eigeneqn})  is equivalent to
\begin{eqnarray*}
 \left[ \beta^{+}_{\pm} \beta_{\pm} + \gamma_+^{+} \gamma_+  + 1 \right] \phi = \frac{\lambda}{4B} \phi,
\end{eqnarray*}
where creation  and annihilation operators  $\beta_\pm$, $\beta_\pm^{+}$, $\gamma_\pm$, $\gamma_\pm^{+}$ 
are expressed in terms of  the  operators
$\alpha^{+}$, $\alpha$:
\begin{eqnarray*}
      \beta_{\pm} = \frac{1}{2}\left( \alpha_1 \mp i \alpha_2 \right),
      \ 
     \gamma_{\pm} = \frac{1}{2}\left( \alpha_3 \mp i \alpha_4 \right),
      \ 
     \alpha_\mu = \frac{1}{\sqrt{B}} (B x_\mu + \partial_\mu),\\
     \beta_\pm^{+} = \frac{1}{2}\left( \alpha^{+}_1 \pm i \alpha^{+}_2 \right),
      \ 
     \gamma_\pm^{+} = \frac{1}{2}\left( \alpha^+_3 \pm i \alpha^+_4 \right),
      \ 
     \alpha^{+}_\mu = \frac{1}{\sqrt{B}} (B x_\mu - \partial_\mu).
\end{eqnarray*}
Here ``$\pm$'' indicates  the self-dual  and anti-self-dual configuration. The eigenvalues  and the square integrable eigenfunctions are
\begin{eqnarray}
\label{efosc}
 && \phi_{n m k l} (x) = \frac{1}{\sqrt{n!m!k!l!}\pi^2} \left(\beta_+^{+}\right)^{k} \left(\beta_-^{+}\right)^{l} \left(\gamma_+^{+}\right)^{n}\left(\gamma_-^{+}\right)^{m} \phi_{0000}
 (x),
 \ \phi_{0000} (x) =  e^{-\frac{1}{2}Bx^2},\\
&& \lambda_{r} = 4B\left( r + 1 \right), 
\nonumber
\end{eqnarray}
where  $r=k+n $ for  self-dual \ field,  $r=l+n$  for  anti-self-dual  field.
The spectrum is discrete. At the domain wall the eigen functions are continuous. There is an infinite degeneracy of the eigen values.

 Inside the domain wall ($x_1=0$)  vector potential can be chosen as  
$$
B_2=0, \ B_1=2Bx_3, \ B_3=0,\  B_4=2Bx_3 \ \ \ (H_i=2B\delta_{i2}, \ E_i=-2B\delta_{i3}).
$$
Charged field displays continuous spectrum  similar to Landau levels.  Square integrable over $x_3$ eigen functions take the form
\begin{eqnarray}
\label{Lpv}
 \phi_n=\exp(-ip_4x_4-ip_2x_2)\chi_n,
\end{eqnarray}
where functions $\chi_n$ 
\begin{eqnarray*}
 \chi_n(p_4|x_3)=\exp\left\{-2\sqrt{2}B\left(x_3+\frac{p_4}{4B}\right)^2\right\}H_n\left(2^{3/4}\sqrt{B}\left(x_3+\frac{p_4}{4B}\right)\right)
\end{eqnarray*}
are solutions of the eigen value problem  
\begin{eqnarray*}
 \left[ p_2^2-\partial_3^2+(p_4+2Bx_3)^2+4B^2x_3^2 \right]\chi_n=\lambda_n \chi_n,\\
\end{eqnarray*}
with the eigen values
\begin{eqnarray*}
\lambda_n(p_2^2,p_4^2)=2\sqrt{2}B\left(2n+1+\frac{p_2^2}{2\sqrt{2}B}+\frac{p_4^2}{4\sqrt{2}B})\right).
\end{eqnarray*}
The character of charged field modes is qualitatively different in the domain bulk (self-dual field) and inside the domain wall (crossed electric and magnetic fields),  which illustrates the character of the problem to be solved to obtain a continous common solution for the domain bulk and wall for the case of the finite width of the kink. The form of the eigen functions (\ref{efosc}) indicates charge field confinement in the bulk and the presence of ``plain wave'' solutions inside the wall.

\section{Conclusions}
In terms of the effective Lagrangian we investigated manifestations of CP and the Weyl group associated with the $SU(3)$ gauge theory. It is shown that the requirement of nonzero gluon condensate leads to the existence of a set of   degenerated  minima and, as a consequence,  triggers the kink-like gauge field configurations interpolating between different minima.  The spectrum of a charged scalar field in the background of  the  kink-like fields was estimated. The bound state form of the eigen functions  (\ref{efosc}) indicates confinement of charged field inside domain, while the ``plain wave'' eigen modes (\ref{Lpv}) exist inside the wall.  The eigenfunction properties and the propagator of a charged field in the kink background have to be studied  in detail for the case of finite width of the kink. 
It is important to investigate the eigenvalue problem for fermionic charged fields and the chiral symmetry realization in the kink-like background. 

The domain model of QCD vacuum developed in~\cite{EN,NK1,NK2} is based on the ensemble of the background gluon  fields with the field strength  being constant almost everywhere in $R^4$. The direction of the field in space and color space as well as duality of the field  are random parameters of the domains. All configurations of this type are summed up in the partition function.  The domain model exhibits confinement of static and dynamic quarks, spontaneous breaking of the flavor chiral symmetry, $U_{\rm A}(1)$ symmetry is broken due to the axial anomaly,  strong $CP$ violation is absent in the model.  The domain boundaries were introduced by means of bag-like boundary conditions imposed on the gluon and quark fluctuation  fields, which made the model unbalanced and considerably complicated all calculations. Gauge field configurations investigated in   the present paper provides us with an interesting option for  parameterization  of the  domain structured ensemble of gluon fields. \\

{\bf Acknowledgments:} We would like to thank G. Efimov, A. Isaev, A. Dorokhov, A. Khvedelidze and N.~Kochelev for discussions and valuable comments.

\addtocontents{toc}{\protect\contentsline{section}{Bibliography}{\thepage}}

\end{document}